\documentclass{cta-author}

{}
{}
{}
\usepackage{multirow}
\begin{document}

\title{A Machine Learning Based Intrusion Detection System for Software Defined 5G Network}

\author{\au{Jiaqi Li}, \au{Zhifeng Zhao}, \au{Rongpeng Li}}

\address{\add{1}{College of Information Science $\&$ Electronic Engineering, Zhejiang University, Zheda Road 38, Hangzhou, Zhejiang Province, 310027, People's Republic of China}
\email{\{21631097, lirongpeng, zhaozf\}@zju.edu.cn}}

\begin{abstract}
As an inevitable trend of future 5G networks, Software Defined architecture has many advantages in providing centralized control and flexible resource management. But it is also confronted with various security challenges and potential threats with emerging services and technologies. As the focus of network security, Intrusion Detection Systems (IDS) are usually deployed separately without collaboration. They are also unable to detect novel attacks with limited intelligent abilities, which are hard to meet the needs of software defined 5G. In this paper, we propose an intelligent intrusion system taking the advances of software defined technology and artificial intelligence based on Software Defined 5G architecture. It flexibly combines security function modules which are adaptively invoked under centralized management and control with a globle view. It can also deal with unknown intrusions by using machine learning algorithms. Evaluation results prove that the intelligent intrusion detection system achieves a better performance.
\end{abstract}

\maketitle

\section{Introduction}\label{sec1}

Software Defined 5G architecture will be a crucial tendency in the development of future 5G networks \cite{1}. It takes the advantage of Software Defined Network (SDN) \cite{2} and Network Functions Virtualization (NFV) \cite{3} through centralized management and dynamic resource allocation to meet the demands of 5G networks. Besides, the separation of the control and execution planes also facilitate the supervision of network status and the collection of information. With the uprising of novel technologies and attacks, it will also be faced with various challenges and severe security situations. As a result, new network security systems and architectures are desperately needed to enhance the security of Software Defined 5G networks \cite{4}.

As an essential technology in network security, intrusion detection systems have received more and more concerns in efficiently detecting malicious attacks. Existing IDS with separate functions are usually deployed locally within restricted areas which are hard to cooperate with each other. Moreover, they are usually signature-based by matching behaviors of incoming intrusions with historical knowledge and predefined rules, which are unable to detect novel attacks intelligently.

To overcome the limitation of traditional IDS, Artificial Intelligence (AI) has been employed for intelligent detection. They classify abnormal traffic using machine learning techniques with a self-learning ability \cite{5}. At present, there have been a few researches in the combinations of IDS and AI. However, they are still inadequate for coordinated detection considering the evolution and development of network systems.

In this paper, we propose an intelligent intrusion detection system for Software Defined 5G networks. Benefit from the Software Defined technology, it integrates relevant security function modules into a unified platform which are dynamically invoked under centralized management and control. Besides, it implements machine learning to intelligently learn rules from huge quantities of data and detects unknown attacks based on flow classification. It uses Random Forest for feature selection and combines k-means++ with Adaboost for flow classification. The proposed system enhances the strength of security protection for future 5G networks.

The remainder of the paper is organized as follows. Section~2 discusses the related research in the field of intrusion detections by machine learning. Section~3 provides an overview of the architecture and describes each module in details. Section~4 introduces the proposed machine learning algorithms for feature selection and traffic classification. Section~5 presents the performance evaluation and results.  Section~6 summarizes the paper and proposes potential future work.

\section{Related work}\label{sec2}

Network intrusion detection has gained intensive discussions and wide investigations in recent years \cite{6} \cite{7}. Most of them depend on a set of manual rules, which are still unable to identify emerging diverse attacks. In terms of this, various machine learning algorithms have been adopted in traffic-based classification in solving such problems  \cite{8} \cite{9}. A hybrid learning approach through combination of K-Means \cite{10} clustering and Decision Tree \cite{11} is proposed in  \cite{12}. Instances are preliminarily separated into different clusters and further classified into more specific catogories. In \cite{13}, the authors propose a three-layer Recurrent Neural Network (RNN) \cite{14} with groups of features as inputs as a neural classifier for misuse detection. In \cite{15}, a systematic framework is presented by applying a data mining technique named Random Forest for misuse and anomaly detection. The performance of the integrated hybrid system is improved through combining the advantages of both the misuse and anomaly detection. However, current IDS monitor the network status through a single point isolately and cannot coordinate with each other.
 
\begin{figure*}[!t]
	\centering{\includegraphics[scale=0.4]{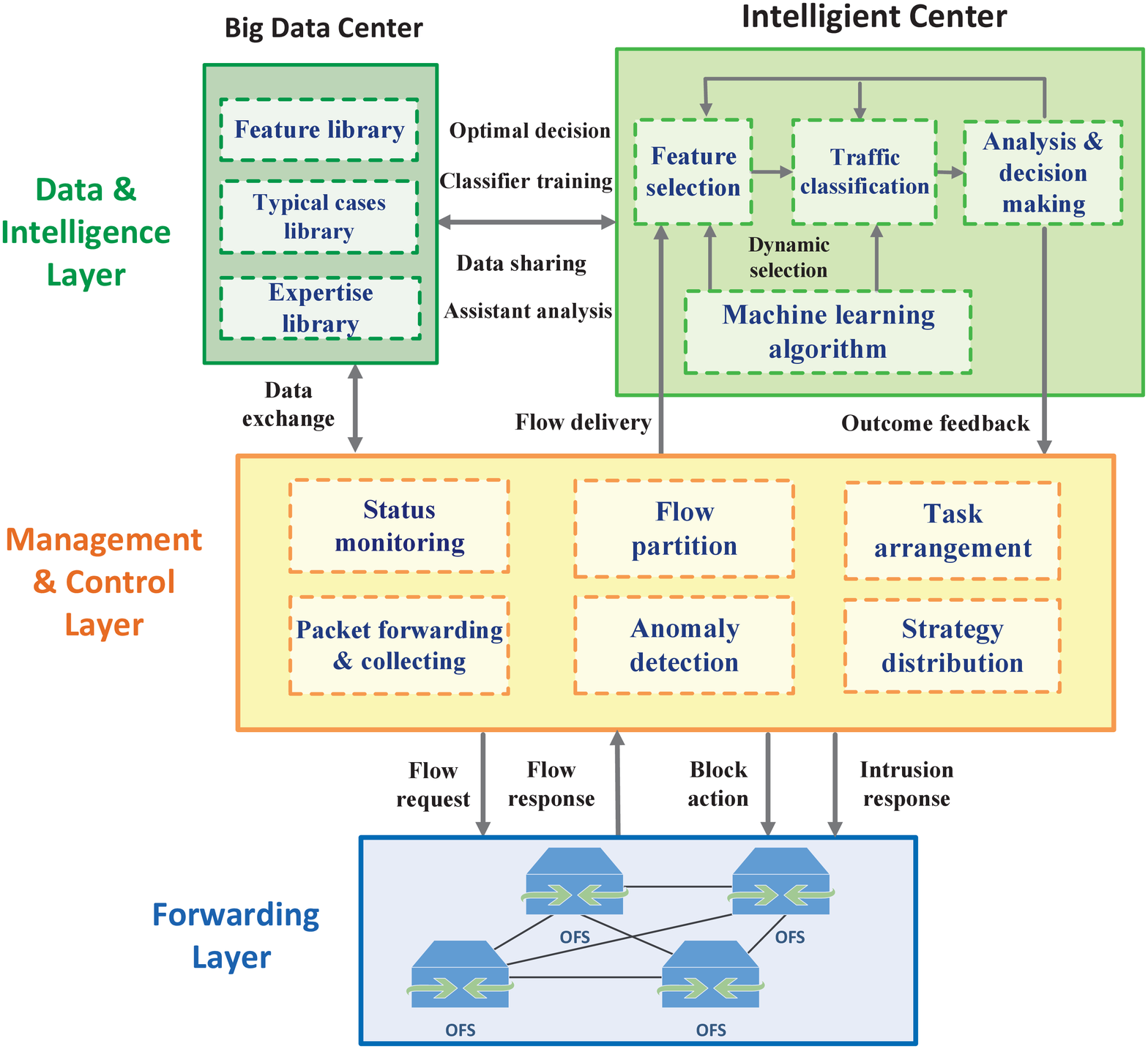}}
	\caption{A machine learning based intrusion detection
		system for software defined 5G network}
\end{figure*}

\section{Architecture}\label{sec3}

The architecture of our proposed intelligent intrusion detection system is illustrated in Figure 1. There are three layers: Forwarding Layer, Management $\&$ Control Layer and Data $\&$ Intelligence Layer. Forwarding Layer consisting of Open Flow controlled entities in 5G (OFs) is responsible for traffic monitoring and capturing. It can collect and upload network flows to the control layer, and block malicious flows according to the instructions of the controller. Management $\&$ Control Layer identifies suspicious flows and detects anomalies preliminarily using uploaded flow information. It also generates protection strategies according to decisions made by the intelligent layer and instructs Forwarding Layer. In Data $\&$ Intelligence Layer, Intelligent Center make further analysis and judgment through feature selection and flow classification using adaptive machine learning algorithms. 

\subsection{Forwarding Layer}\label{subsec3.1}

This layer is in charge of forwarding packets between OFs. It provides Management $\&$ Control Layer with real-time network status through collecting and uploading anomaly information from distributed OFSs. Besides, intrusions can also be blocked by OFSs through dropping malicious packects under the command of upper layers.

\subsection{Management and Control Layer}\label{subsec3.2}

\subsubsection{Packet collecting $\&$ Flow partitioning}\label{subsubsec3.2.1} 

This layer provides a more global view of the entire 5G network. The status monitoring module supervises network status and periodically requests packets from OFs. It collects packets uploaded by OFs for further analysis. Flow partition module processes and parses traffic statistics, clusters packets into flows and generates 6-tuple Flow IDs as follow.
\begin{equation}
\label{eq:flow profile}
\begin{aligned}
&\{\textit{srcip,dstip,srcport,dstport,duration,protocol}\}
\end{aligned}
\end{equation}
The Flow IDs are used to define and label different flow records representing specific network connections and activities. The packet collection and inspection is performed at a regular time interval. The time interval is delicately selected in order to avoid undesirable delays for anomaly identification and scale packet overhead to an acceptable level.


\subsubsection{Anomaly detection}\label{subsubsec3.2.2}

Before exploiting elaborate intrusion detections in the intelligence layer, some basic flow statistics are used to roughly recognize abnormal behaviors and potential anomalies. The module applies entropy analysis \cite{16} based on Shannon Theory to detect distribution variations of packet sequences. The entropy of a random variable $x$ is:
\begin{equation}
\label{eq:entropye}
\begin{aligned}
&H(x) =  - \sum\limits_{i = 1}^N {p(x_i)} \log (p(x_i))
\end{aligned}
\end{equation}
where $x_i$ is the value of $x$ ranging from 1 to $N$. $p(x_{i})$ calculates the possibility of $x$ being  $x_{i}$  observed among all the feasible values.
Here we consider four basic characteristics: source address, source port, destination address, destination port as variables in the above Eq.(2) which are sketched from packets in every consecutive duration. Within the given period of time, we compute a new entropy $H(x)$ of each characteristic to detect anomalies in the subsequent way. If $E$ stands for the mean entropy and $S$ represents the corresponding standard deviation, there will be possible anomalies if $H(x)$ falls outside the duration between ($E$-$S$) and ($E$+$S$). The suspicious groups of flows are delivered to the Intelligent Layer for further analysis \cite{17}.

\subsubsection{Task arrangement $\&$ Strategy distribution}\label{subsubsec3.2.3}

The modules response to detected intrusions through managing and organizing specific  actions for defending attacks. They develop optimal strategies, arrange objective tasks and distribute them to OFs in Forwarding Layer. OFs are instructed to drop packets of malicious flows and protect the 5G network from being further attack.

\subsection{Data and Intelligence Layer}\label{subsec3.3}

\subsubsection{Feature selection}\label{subsubsec3.3.1}

The feature selection module is designed to extract concerned features of skeptical flows and find an optimal subset of preferable features. They are used to precisely describe and discriminate a flow. The module can process high dimension data efficiently and remove irrelevant data, which improves the learning efficiency and predictive accuracy of flow classification. The selected features are considered optimal if they are closely correlated to the correct classification result while not redundant. In our system, various algorithms can be selected to measure the relevance and redundancy of features.

\subsubsection{Traffic classification}\label{subsubsec3.3.2}

The module classifies network flows by marking whether it belongs to specific types of attacks or benign traffic. The output of the classifier labels each flow as a certain class. Combination algorithms can be used to increase the accuracy for machine learning classification. 

\subsubsection{Analysis $\&$ decision making}\label{subsubsec3.3.3}

According to the classification results, the module makes a comprehensive analysis of the real-time network status and determines whether the network is under attack or not. The analysis results can be feedback to the former two modules and assists them in selecting algorithms adaptively. More important, they are delivered to the controller for tactics arrangement and defense.

\subsubsection{Big data center}\label{subsubsec3.3.4}

As an auxiliary module in Data $\&$ Intelligence Layer, Big Data Center maintains various bases of historical records and knowledge of intrusions to facilitate classifier training and decision making. It is comprised of libraries of typical flow features, user activity models and expertise advice. The data is persistently revised and updated.

\section{Intelligent Intrusion detection process}\label{sec4}

In our system, we employ selected machine learning methods in two critical steps of intrusion detection. Firstly, we use Random Forest (RF) to select optimal subset of flow features through measuring variable importance. Afterwards, a hybrid clustering-based Adaboost will classify traffic into different classes of attacks with selected features as input. 

\subsection{Random Forest}

Random Forest \cite{18} is a collection of uncorrelated structured decision trees deemed as forest. Those trees make classifying judgments independently and the final result will be the one gaining the majority votes. The concept of `random' typically manifests in two aspects as follows:

$\bullet$ If the number of input training data is N, we take N samples randomly with replacement  from the original data. The selected samples are likely to be repeated and will be used for growing the tree. Meanwhile, those data which has not been selected to build a tree is known as out-of-bag (OOB) data. The data is utilized to measure the classification accuracy of the forest. 

$\bullet$ For each tree, we choose m ($m < M$, usually $m = \sqrt M $) features out of M-attribute entire set randomly as input variables without replacement. The value of m remains unchanged during the whole process of forest construction. While determining the best feature at each splitting node, we calculate Gini Ratio \cite{19} of m attributes and choose the one with the highest value to split the node. We will stop growing the tree when the selected attribute is the same as its father node. 

The algorithm is a valid ensemble machine learning algorithm used for classification and regression. There is no need to prune each tree as it grows in case of over-fitting under those two restrictions. In particular, it can also be used to select features by ranking the significance of different features \cite{20}. The measuring of  variable importance is based on classification accuracy of out-of-bag data. The main steps are showed in Figure 2.

\begin{figure}[!b]
	\centering{\includegraphics[scale=0.31]{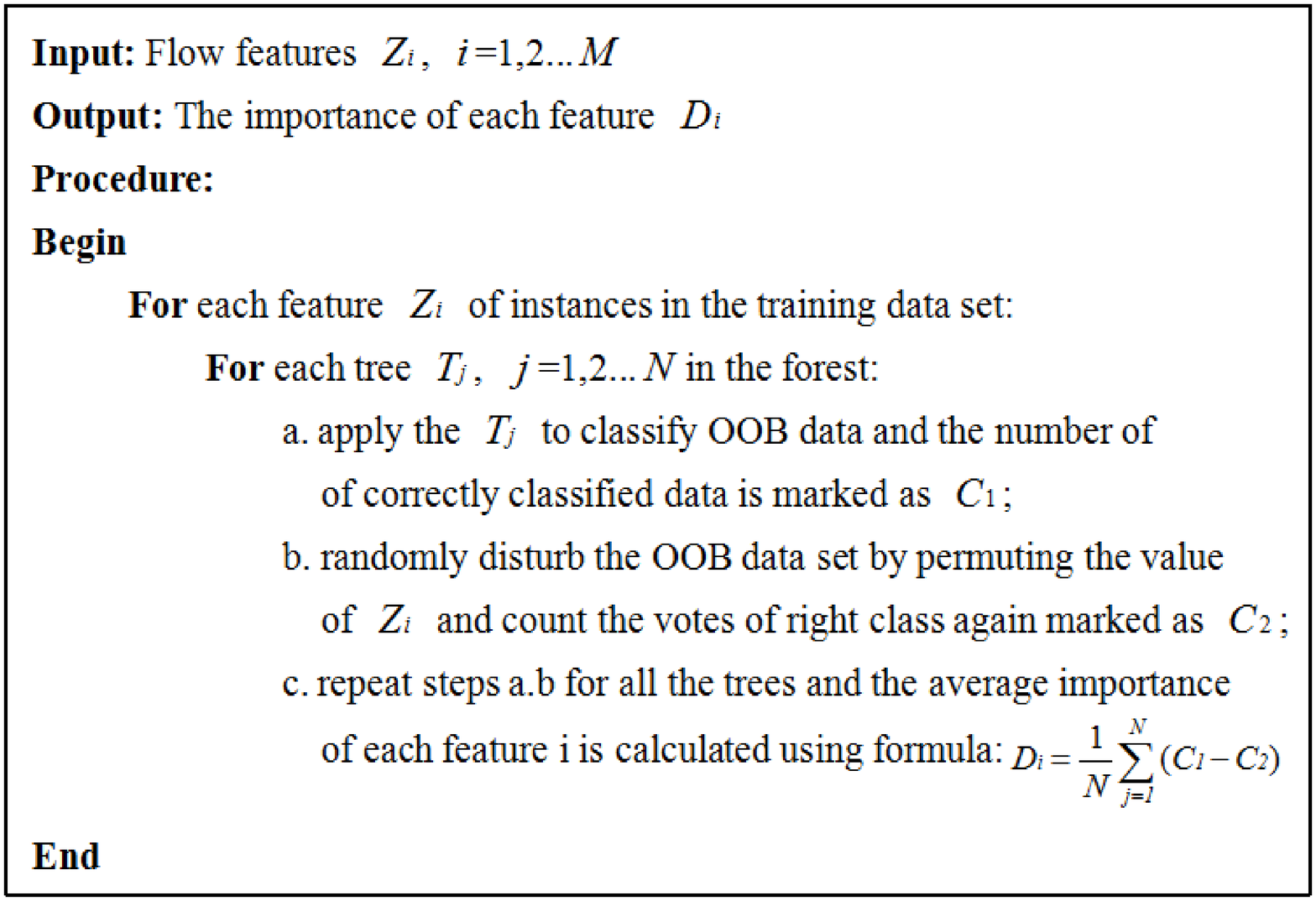}}
	\caption{Importance measurement of flow features using Random Forest\label{fig1}}
\end{figure}

We obtain the variable importance of the M features and select those with higher values in accordance with the pre-established number of features or retaining ratio. The importance of each feature is measured by the influences it exerts on the result of classification. A specific feature with higher importance usually degrades the out-of-bag accuracy to a great extent.

\subsection{Hybrid Clustering-Based Adaboost}

There are two steps in traffic classification. For the first stage, we make a preliminary judgment by adopting k-means++ to divide the traffic into two clusters which most probably represent the normal and abnormal instances. Later, we further partition the anomaly clusters into four main classes of attacks using the ensemble algorithm Adaptive Boosting (Adaboost).

\begin{figure}[!t]
	\centering{\includegraphics[scale=0.28]{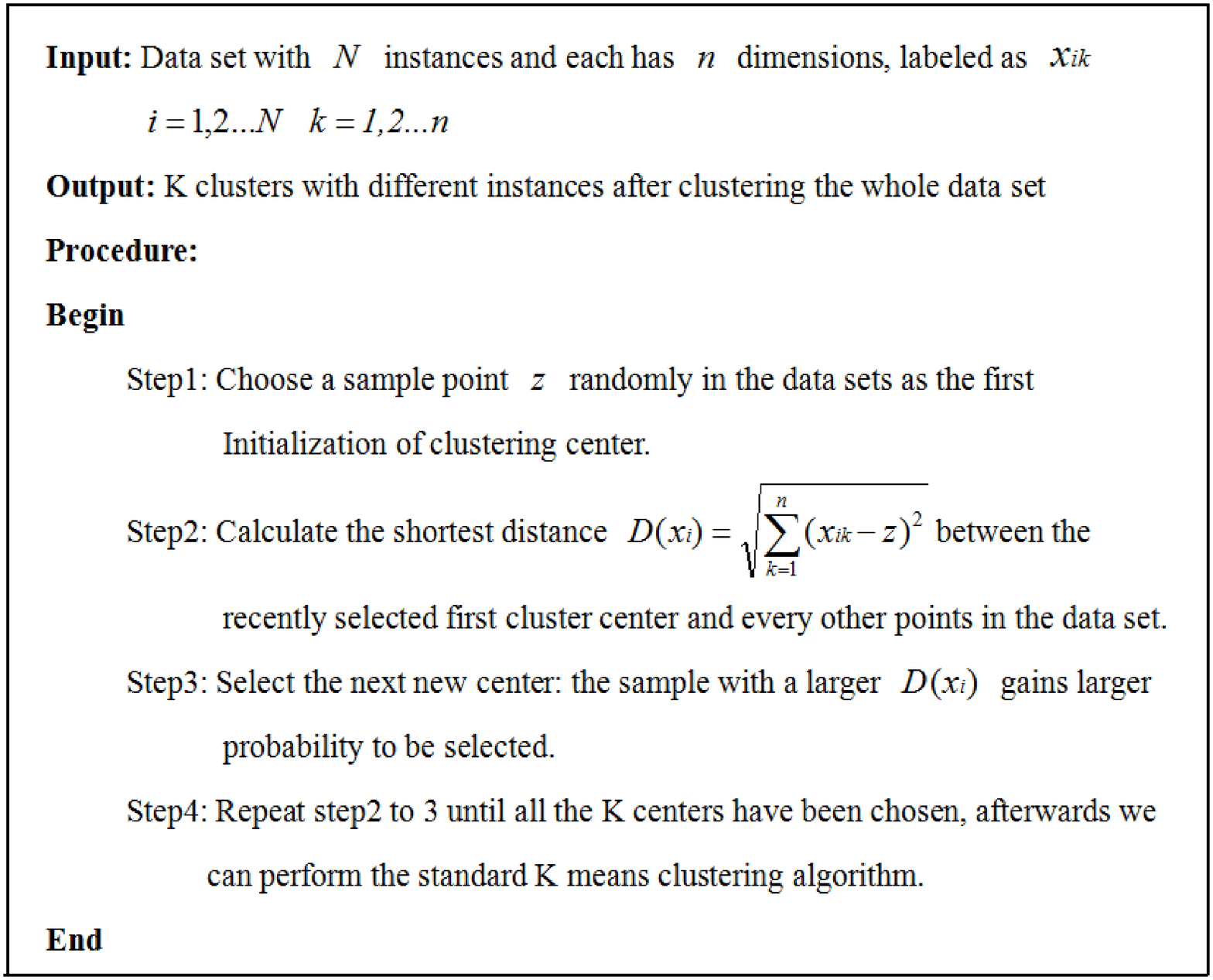}}
	\caption{Main steps of k-means++ algorithm\label{fig1}}
\end{figure}

\subsubsection{k-means++}

As an unsupervised learning algorithm, the dominate object of applying k-means clustering method is to separate and group unlabeled traffic into normal and attack classes for coarse-grained classification. However, it is worth noting that the number of centroids k requires to be pre-determined and the random selection of initial centroids may result in locally optimal clustering. Hence, we introduce an improved k-means++ technique \cite{21} aiming at choosing the optimal clustering centers. The fundamental principle we comply with is to make the distances between any initial clustering centers as far as possible. The process of initialing centroids is described in Figure 3.

For step3, the key point is that how to reflect the relationship between the distance variable $D(x_{i})$ of one sample and its possibility of being selected. We aggregate the distance $D(x_{i})$ of each point into a whole assemblage in sequence and compute the sum of them as $Sum(D(x_i)) = D(x_1) + D(x_2) + ...D(x_N)$. Then a random value $\lambda$ is selected which is located in the range of 0 to $Sum(D(x_i))$. We update the value of $\lambda$ by computing $\lambda=\lambda-D(x_i)$ from i=1 to N until it falls below zero. At this time, the section which  $\lambda$ drops in with the length $D(x_{i})$ indicates the corresponding sample point $i$ to be the next centroid. Now that the value of $\lambda$ is stochastic, there is more chance that it drops in the section of a larger $D(x_{i})$. In this paper, we set K=3, considering that two of the four kinds of attacks (U2R, R2L) are easy confused with the normal flows.

\subsubsection{Adaboost}

Adaboost is a strong ensemble classifier linearly composed of different weak classifiers after trained by the same set of data \cite{22}. 

The weight of each sample is same while initializing the training data. At each iteration, the data distribution is adaptively altered through changing the weight of samples. The weight of the sample which is mis-classified in the former basic classifier will be improved in the next round of training. In contrast, it will be reduced if it is correctly classified. In this way, more attention is laid on the samples hard to be properly classified to promote the overall performance.

All the weak classifiers after training are combined assigned with different weights of contribution and form a strong classifier. Thus, the final result is determined by votes of each basic classifiers with distinct right of speech $\alpha$. $\alpha$ is inversely proportional to the classification error rate $e$, which indicates that those weaker classifiers gaining a higher classification accuracy contribute more to the final result. Each weak classifier is regulated by factor $\alpha$ and the linear formation of them achieves a better result.

In most cases, each weak classifier is constructed by a one-layer decision stump, which splits once solely based on a single feature. It is worth noting that the feature used at each decision tree for decision-making in classification is optimally chosen from N features. The feature used at each classifier is totally independent and can be reused again.

\section{Experiment result}
\label{sec:evaluation}

In this section, we conduct several experiments to evaluate our proposed system.

\subsection{Dataset}

The KDD Cup 1999 dataset has been widely used to evaluate the performance of intrusion detection methodologies in recent years \cite{23}. It contains approximately 5,000,000 network connections in the training set and nearly 2,000,000 instances in the testing set. Each single connection vector consists of 41 features sorted into three classes: basic connection-based feature, content-based feature, traffic-based feature. Each traffic sample is labeled as either a normal flow or a malicious intrusion which exactly falls into 4 different categories in accordance with their own characteristics: DoS (Denial of Service), R2L (Remote-to-local), U2R (User-to-Root) and Probe. Since the amount of the original dataset is huge, we perform a five-class flow classification emulation using 10$\%$ of the whole KDD99 intrusion detection raw dataset. The distribution of both training and testing data marked by their attack type is summarized in Table \uppercase\expandafter{\romannumeral1}.
\begin{table}[!t]
	\processtable{Distribution of data used in our evaluation\label{tab1}}
	{\begin{tabular*}{20pc}{@{\extracolsep{\fill}}lllllll}\toprule
			\multirow{2}{*}{Class} & \multicolumn{2}{c}{Training dataset} & \multicolumn{2}{c}{Testing dataset} \\ 
			& No. of samples      & Percentage      & No. of samples      & Percentage     \\\midrule
			Normal                 & 97278               & 19.69\%         & 60593               & 19.48\%        \\ 
			Probe                  & 4107                & 0.83\%          & 4166                & 1.34\%         \\
			DoS                    & 391458              & 79.24\%         & 229853              & 73.9\%         \\
			U2R                    & 52                  & 0.01\%          & 228                 & 0.07\%         \\ 
			R2L                    & 1126                & 0.23\%          & 16189               & 5.2\%          \\\botrule
		\end{tabular*}}{}
	\end{table}
	
\subsection{Evaluation Metrics}\label{subsec6.1}

Generally, the performance of the intrusion detection system is evaluated in the light of precision (P), recall (R), F-score (F), accuracy (AC) and false alarm rate (FA) calculated in the formulas below. We desire a system with higher detection rate as well as lower false rate.\\
Precision (P): the percentage of intrusion predicted that is truly existed. 
\begin{equation}
\label{eq:precision}
\begin{aligned}
&P = \frac{{TP}}{{TP + FP}}
\end{aligned}
\end{equation}
Recall (R): the number of correctly predicted intrusions versus all the presenting intrusions.
\begin{equation}
\label{eq:recall}
\begin{aligned}
&R = \frac{{TP}}{{TP + FN}} 
\end{aligned}
\end{equation}
F-score (F): makes a tradeoff between the precision (P) and recall (R) to reach a better measurement of classification accuracy.
\begin{equation}
\label{eq:F-score}
\begin{aligned}
&F = \frac{2}{{\frac{1}{P} + \frac{1}{R}}}
\end{aligned}
\end{equation}
Accuracy (AC): manifests the flows exactly classified over the entire traffic traces.
\begin{equation}
\label{eq:accuracy}
\begin{aligned}
&AC = \frac{{TP + TN}}{{TP + TN + FP + FN}}
\end{aligned}
\end{equation}
False positive rate (FPR): indicates the percentage of normal traffic which is mis-classified as attacks.
\begin{equation}
\label{eq:accuracy}
\begin{aligned}
&FP = \frac{{FP}}{{FP + TN}}
\end{aligned}
\end{equation}
Where:\\
TP: the number of attacks precisely detected.\\
TN: the number of normal traffic precisely classified.\\
FP: the number of normal traffic incorrectly classified.\\
FN: the number of attacks unsuccessfully detected.

To our common sense, various intrusions generate different level of consequences to the entire network. Hence, another comparative metric is defined to measure the cost damage of misclassification for different attacks per sample calculated as below:
\begin{equation}
\label{eq:cost}
\begin{aligned}
&Cost = \frac{1}{N}\sum\limits_{i = 1}^n {\sum\limits_{j = 1}^n {Mij \times Cij} } 
\end{aligned}
\end{equation}
where $M_{ij}$  indicates the number of instances of class $i$  misclassified in class $j$. $C_{ij}$ the cost value representing the penalty for each sample obtained from cost matrix \cite{24} employed for KDD99 shown in Table \uppercase\expandafter{\romannumeral2}. Let $N$ be the total number of samples for testing.
\begin{table}[!t]
	\processtable{Cost matrix\label{tab2}}
	{\begin{tabular*}{20pc}{@{\extracolsep{\fill}}llllll}\toprule 
			Class  & Normal & Probe & DoS &U2R & R2L \\ \midrule
			Normal &0     & 1     & 2  & 2 & 2   \\ 
			Probe  & 1      & 0     & 2   & 2   & 2   \\ 
			DoS    & 2      & 1     & 0   & 2   & 2   \\ 
			U2R   & 3      & 2     & 2   & 0   & 2   \\ 
			R2L    & 4      & 2     & 2   & 2   & 0   \\ \botrule
		\end{tabular*}}{}
	\end{table}

\subsection{Performance Analysis}

The performance of detection using the sub-feature dataset selected by Random Forest in contrast with a full-feature dataset is demonstrated in Table \uppercase\expandafter{\romannumeral3}. It is encouragingly illustrated that the selected features contribute more to differentiate attack traffic giving rise to higher accuracy and lower false rate. As noticed from the figure, the selected features achieve a slight reduction of operating time as well as lessening the computational cost without degrading the overall performance.

\begin{table}[!b]
	\processtable{Performance comparision using differnet number of features\label{tab3}}
	{\begin{tabular*}{21pc}{@{\extracolsep{\fill}}llllllll}\toprule 
			Number of  & Precision & Recall  & F$\_$score & FPR & Cost & Time\\ 
			feature   &($\%$)  &($\%$)  &($\%$)  &($\%$) & &(Seconds)\\\midrule
			23                 & 94.48          & 92.62       & 91.02         & 0.54     & 0.2410 & 110           \\
			41                 & 93.60          & 92.06       & 90.03         & 0.54     & 0.2574 & 149           \\ \botrule
		\end{tabular*}}{}
	\end{table}
\begin{figure}[!b]
	\centering{\includegraphics[scale=0.5]{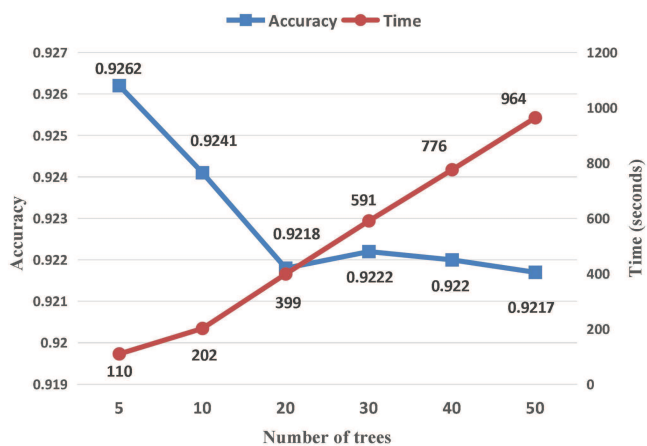}}
	\caption{Performance with different values of parameter $NT$ }
\end{figure}
	
To acquire a better performance, we try to prune the basic parameter of Adaboost by building the ensemble classifier with different numbers of weak estimators. It means the number of trees (NT) built as basic learners. The accuracy rate and time for running the algorithm are plotted in Figure 4. We strike a balance between the two factors and choose NT=5 as the desirable value in the following evaluation experiments. It is preferable that the system can reach the similar performance with effectively less computational consumption avoiding high resource utilization.

\begin{figure}[!t]
	\centering{\includegraphics[scale=0.5]{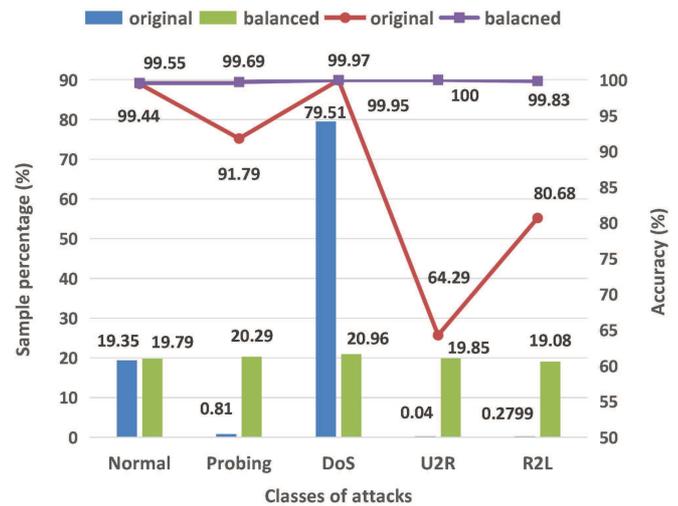}}
	\caption{Accuracy on the balanced dataset compared with the original dataset}
\end{figure}

In the real network, some intrusions generate more connections than others which lead to an extreme unbalanced dataset for classification. Thus, we resolve the problem through down-sampling the majority intrusions (normal and DoS) as well as oversampling the minority intrusions (U2R and R2L). We make a comparison of detection rate between the ordinary dataset and the balanced dataset using the same default parameters by splitting 40$\%$ of the complete set as testing data. The result is demonstrated in Figure 5. It gives an overview of percentages of samples classified into five classes and the distributions show extreme variations between the two dataset. It is apparent that the sampling technique improving the detection accuracy of minority intrusions dramatically while maintains a reasonable detection rate of the majority ones. The sampling result indicates that pre-processing the input data into uniform distribution upgrades the detection of minority intrusions which elevates the overall performance of our system.

\begin{table}[!b]
	\processtable{Performance comparision using differnet number of features\label{tab3}}
	{\begin{tabular*}{20pc}{@{\extracolsep{\fill}}llllllll}\toprule 
			\multicolumn{2}{l}{Combination of}  & No. of & Precision & Recall  & F$\_$score & FPR \\ 
			\multicolumn{2}{l}{algorithm}   & features  &($\%$)  &($\%$)  &($\%$) &($\%$) \\\midrule
			RF                               & KA                            & 23                               & 94.48                          & 92.62                       & 91.02                         & 0.54                     \\ 
			RF                               & GBDT                          & 23                               & 93.09                          & 91.21                       & 89.37                         & 2.84                     \\ 
			RF                               & DT                            & 23                               & 92.65                          & 91.78                       & 90.01                         & 3.31                     \\ 
			RF                               & SVM                           & 23                               & 90.14                          & 91.46                       & 89.44                         & 1.47                     \\ 
			Tree                             & KA                            & 23                               & 93.34                          & 91.90                       & 89.99                         & 0.64                     \\ 
			Fisher                           & KA                            & 10                               & 93.25                          & 91.72                       & 89.79                         & 1.91                     \\ 
			ReliefF                              & KA                            & 8                                & 91.55                          & 90.96                       & 89.07                         & 8.35                     \\ \botrule
		\end{tabular*}}{}
	\end{table}
	
\begin{table}[!b]
	\processtable{Classification accuracy ($\%$) comparision between normal process and processes with CV in each type of attack\label{tab3}}
	{\begin{tabular*}{20pc}{@{\extracolsep{\fill}}lllllll}\toprule 
			 & Normal & Probe & DoS   & U2R  & R2L \\\midrule
			Normal process without CV    & 99.46  & 73.89 & 97.36 & 0.88 & 5.8   \\
			Training data with CV & 99.9   & 97.04 & 97.04 & 12.5 & 88.46 \\ 
			Testing data with CV & 98.54  & 97.96 & 99.97 & 68   & 65.5 \\ \botrule
		\end{tabular*}}{}
	\end{table}
	
Since we know that the selection of algorithms for feature selection and traffic classification have a mutual influence on each other, we care more about the performance of the combination of them. Here several groups of traditional feature selection and machine learning algorithms are served as comparisons in Table \uppercase\expandafter{\romannumeral4}. As we can see, it is obvious that our methods generate a better performance among all the combination alternatives in every metric. It is desperate for coupling the selection of algorithms in these two main steps of our system in the long run.

Finally, we verify the classification accuracy of our methodology through cross validation (CV) \cite{25} by splitting 90$\%$ of the dataset for training while the rest 10$\%$ for testing in Table \uppercase\expandafter{\romannumeral5}. It is clearly depicted that classifications with cross validation behave well by significantly boost the accuracy of detecting U2R and R2L attacks as well as gently promoting the rate of other categories. To our knowledge, the four main types of attacks mentioned above are subdivided into 39 small classes, 17 of which only appear in the model we come up with testing data. So it shows that our proposed classifier lacks a generalization ability to detect various attacks without previous training.

\section{Conclusion}
\label{sec:conclusion}
This paper presents an intelligent intrusion detection system based on Software Defined 5G architecture using machine learning algorithms. It is implemented under Software Defined environment which facilitates status monitoring as well as traffic capturing under a global view. It integrates and coordinates security function modules and detects intrusions intelligently based on flow classification. We use Random Forest to select a subset of typical traffic features and classify network flows by combining k-mean++ and Adaboost algorithms. Evaluation results validate the effectiveness of our proposed system in detecting network intrusions.

In the future, we intend to find the intrinsic relations between features as well as classifiers and adaptively choose the best combination of learning approaches.


\section{Acknowledgments}

This work was supported in part by the Program for Zhejiang Leading Team of Science and Technology Innovation (No. 2013TD20), in part by the National Postdoctoral Program for Innovative Talents of China (No. BX201600133), and in part by the Project funded by China Postdoctoral Science Foundation (No. 2017M610369).

\end{document}